\documentclass[sigconf]{acmart} 
\usepackage[utf8]{inputenc}
\usepackage{graphicx}
\usepackage{cleveref}
\usepackage{soul}
\usepackage{color}
\usepackage[caption=false]{subfig}

\title{Building Implicit Vector Representations of Individual Coding Style}

\author{Vladimir Kovalenko}
\affiliation{%
  \institution{JetBrains Research}
  \city{Amsterdam}
  \country{The Netherlands}}
\email{vladimir.kovalenko@jetbrains.com}

\author{Egor Bogomolov}
\affiliation{%
  \institution{JetBrains Research}
  \city{Saint Petersburg}
  \country{Russia}
}
\email{egor.bogomolov@jetbrains.com}

\author{Timofey Bryksin}
\affiliation{%
  \institution{JetBrains Research}
  \city{Saint Petersburg}
  \country{Russia}
}
\email{timofey.bryksin@jetbrains.com}

\author{Alberto Bacchelli}
\affiliation{%
  \institution{University of Zurich}
  \city{Zurich}
  \country{Switzerland}
}
\email{bacchelli@ifi.uzh.ch}

\begin{document}

\begin{abstract}
With the goal of facilitating team collaboration, we propose a new approach to building vector representations of individual developers by capturing their individual contribution style, or \textit{coding style}.
Such representations can find use in the next generation of software development team collaboration tools, for example by enabling the tools to track knowledge transfer in teams.
The key idea of our approach is to avoid using explicitly defined metrics of coding style and instead build the representations through training a model for authorship recognition and extracting the representations of individual developers from the trained model.
By empirically evaluating the output of our approach, we find that implicitly built individual representations reflect some properties of team structure: developers who report learning from each other are represented closer to each other.
\end{abstract}

\maketitle

\section{Introduction}
Machine learning (ML) has lately been having more impact in software engineering by improving over state of the art in problems such as code summarization~\cite{haiduc2010supporting, allamanis2016convolutional} and program synthesis~\cite{devlin2017robustfill}.
Many of the methods that apply ML techniques to code are aimed at enhancing software development tools by offering engineers assistance in routine tasks. Examples of such enhancements include code completion engines, static analysis, and automated code review systems~\cite{raychev2014code}.
Most of these methods are designed to assist in a task that is relevant within a short time scope, such as to insert the right code snippet or to fix problems in a single changeset. 

While this assistance to developers promises a significant improvement in the daily experience with developer tools, it does not cover the complete scope of potential tooling support for engineering teams, particularly when in relation to socio-technical aspects.
Indeed, developers report that their use of developer tools is not only related to technical artifacts, but is also a vital part of interpersonal communication in teams (e.g., as it is in the case of code review tools~\citep{bacchelli2013expectations}).

Social aspects in teams are manifested in technical artifacts and records of these processes can be extracted with software repository mining techniques~\citep{valetto2007using}.
Still, modern software team collaboration tools make little to no use of the data available in software repositories to assist with social aspects of the engineering process. 
To enable tools to assist with interpersonal processes at the scale of a software team, one first needs reliable and transparent models of these processes as well as methods to retrieve corresponding data from software repositories. As a step in this direction, in this work, we propose a new approach to building representations of developers' individual coding fingerprints -- or their \emph{coding style}.
Such representations can find use in the next generation of team collaboration tools, which could, for example, track the process of knowledge transfer in teams and provide assistance. Other potential applications include searching for similar developers and profiling of individual coding habits for tasks related to the management of human resources.

Existing work on code stylometry typically relies on explicitly defined features to represent code style~\cite{caliskan2015deanonymizing}. 
We take a different direction: 
Instead of using explicit measures of code style, we implicitly extract the distinguishing features of individual developers by training a model to recognize authorship of a batch of code changes and processing the model's internal representations. The input of the model consists of changes made by a developer to individual methods, and its output is a label for the predicted author. 
To maximize transparency of the model, we use an attention mechanism -- a technique widely used in neural machine translation~\cite{bahdanau2014neural}, which allows us to point out particular code constructs that are more important for authorship attribution.
After training the model to recognize the author of code changes, we extract representations of contributions of individual developers from it. We do so by combining the vector representations of individual method changes made by a developer over a time period, using the weights for vectors of individual method changes that are learned by the attention mechanism.

Finally, we assess the capability of representations of reflecting a practical social aspect, namely, learning within teams. For this, we produce multiple snapshots of individual representations of developers in a large open source project maintained by an enterprise, with each snapshot corresponding to a specific time period. Thanks to a localized development team, we were able to also collect reports of mutual learning from the developers of this project. 
Finally, we look for a connection between reported learning and relative distances between developer representations.
While we find no connection between reported learning and relative movement of developers' representations between consecutive time buckets, we see that reported learning is associated with lower distance between representations of two developers.

The primary contribution of this work is a novel method to extract representations of the contribution style of individual developers. The method is designed to be suitable for use with raw data from software repositories and to not require any additional labeling or explicit feature engineering. Another contribution is an empirical assessment of how the retrieved representations match learning as perceived by software developers.

\section{Background and motivation}

\subsection{The need for developer representations}\label{sec:representations-need}

In the following, we reason about the importance of representations of individual developers' style, focusing on another class of tools (i.e., IDEs) as an example of a class of advanced tools successfully making use of a comprehensive model of the main medium they are designed to manipulate: code.

Despite the important role of team collaboration tools in software engineering, existing approaches aimed at improving software engineering tools with data have been mostly targeting coding environments and IDEs. 
Modern industry-grade IDEs, such as IntelliJ IDEA~\cite{intellij-idea} and Eclipse~\cite{eclipse-ide}, provide rich toolkits for code manipulation and maintenance. 
The IDEs feature automated code refactoring, code inspections pointing at potential issues, are able to provide high-level overview of large codebases, and enable deep integration of external tools, e.g. debuggers, with the code editor. 
Capability of IDEs to provide such rich code manipulation features is based on comprehensive internal models of software projects. In particular, IntelliJ relies on PSI~\cite{intellij-psi} that represents a rich internal code model. Language-specific features in Eclipse IDEs are based on language support packages like JDT~\cite{eclipse-jdt} and CDT~\cite{eclipse-cdt} that as well manipulate with comprehensive language-specific program structure data.
Outside the industrial IDE realm, academic methods aimed at coding manipulation and improvement also operate with code models and in some cases partially rely on language support initially designed for use in IDEs ~\cite{DBLP:conf/kbse/FalleriMBMM14}.

Modern team collaboration tools, such as code review tools, repository hosting engines, and bug tracking systems, are vital mediums for collaborative software engineering. In fact, these tools do not simply provide an environment to perform short-term tasks, like reviewing changes or communicating an issue, but also play a crucial role in supporting knowledge transfer in teams~\cite{bacchelli2013expectations} and serve as a knowledge base~\cite{tran2008crawling}.

In contrast to comprehensive code manipulation and problem detection features in IDEs, most team collaboration tools, despite their vital role in team-wide processes, do not maintain a comparably complex and detailed model of teams' communication structure, nor do these tools routinely analyze records of prior communication in teams. While there are exceptions to this rule, such as data-driven techniques like reviewer or assignee recommendation systems~\cite{thongtanunam2015should, kovalenko2018does, anvik2006should} and repository analytics features that are present in some collaboration tools, these tools are yet to evolve to feature and utilize a more comprehensive model of team communication and to assist in maintenance and improvement of communication at a larger time scale.

Enabling assistive features in team collaboration tools requires enabling tools to model social processes internally. While existing research suggests that social processes are to an extent reflected in technical artifacts~\cite{cataldo2008socio} and can be extracted with data mining techniques~\cite{valetto2007using}, it is important to focus on extracting representations of individual properties of contributors from records of their collaborative work.

This work is dedicated to extracting representations of individual properties of engineer's coding that distinguish their contributions from their peers'. This could provide the tools with a sense of proximity of individual properties of their users' work, which could be use to detect learning in teams or provide onboarding assistance. We require that extraction of representations should not rely on any explicitly defined set of features, as opposed to existing code stylometry approaches. 
Using a neural code change embedding technique, we avoid feature engineering and utilize the ability of the model to capture optimal distinguishing features implicitly.

\subsection{Existing work}
Despite a solid track record of academic efforts and lack of widespread usage in modern team collaboration tools, we believe individual developer representations to be a promising ground for the evolution of collaboration tools.

The idea of building representations of individual developers' style has been around in the research community for several decades. The need for such representations is mostly motivated by the demand for code authorship attribution, which is deemed important for a variety of real-world applications, such as malware detection~\cite{caliskan2015deanonymizing} and plagiarism elimination~\cite{lange2007using}.

Some recent work is closely related to ours. \citet{azcona2019user2code2vec} propose building vector representations for individual computer science students, based on source code of their assignment submissions. As opposed to our work, they do not use any information on structure of code. 
\citet{alsulami2017source} use a deep learning model to attribute authorship of source code, based on traversal sequences of the AST. Authorship attribution, however, is the sole task of their approach. Moreover, the model they propose works on code snippets and not code modifications, thus making it very hard to apply to data from software repositories.

\section{Method}\label{sec:method}
\begin{figure}[htp]
\includegraphics[clip,width=\columnwidth]{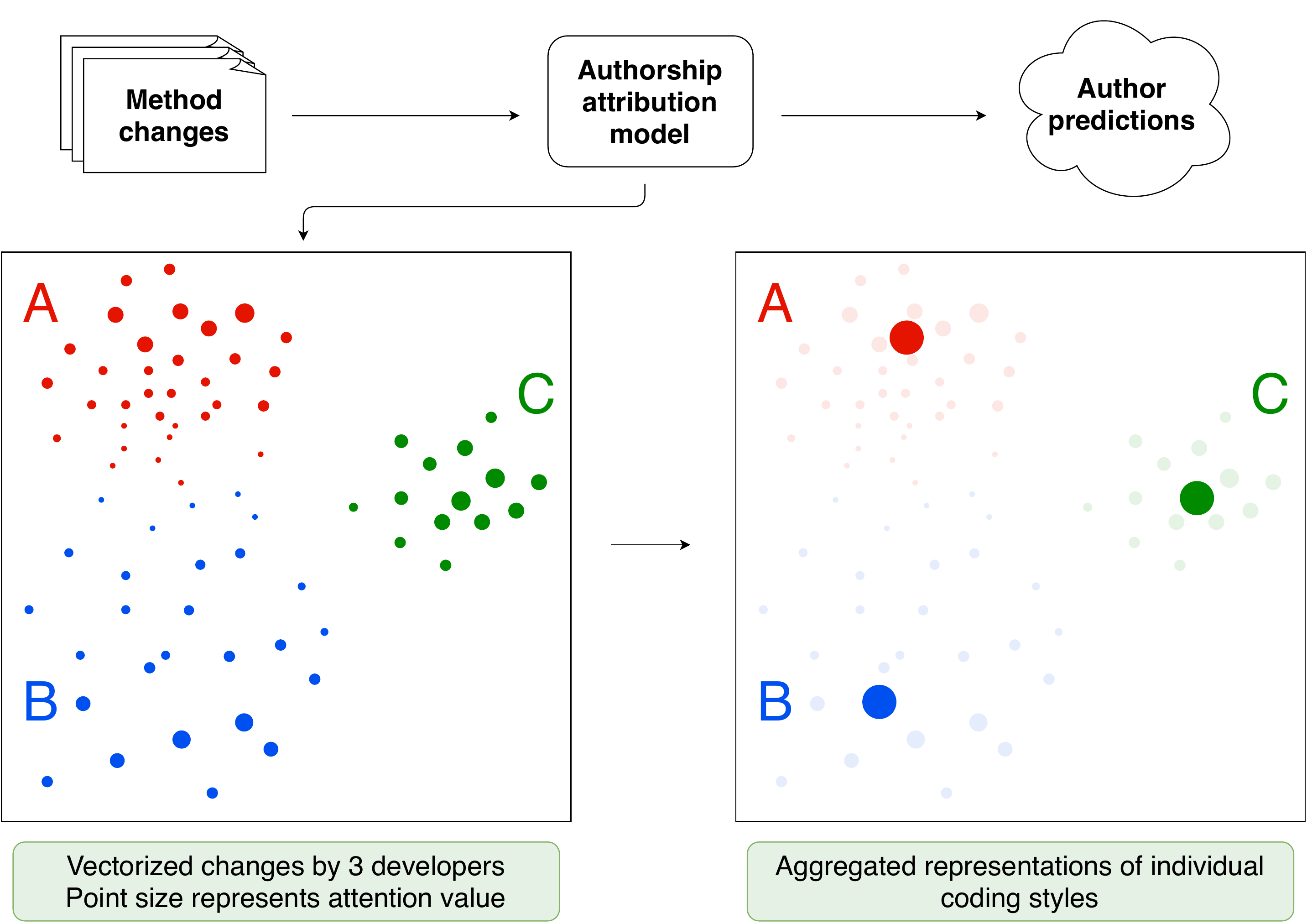}%
\caption{A high-level overview of our approach to building representations of individual developers}
\label{fig:method-hl}
\end{figure}

In contrast to explicitly defined feature sets for developers' coding style, commonly used in existing literature, we define individual coding style in the scope of a single repository as vaguely as \emph{anything that distinguishes a developer's contribution to the codebase from their peers' contributions} and focus our method on capturing this.

We propose a two-step method for building the embeddings -- essentially, vector representations -- of individual code style. 
The overarching idea of our approach is to first learn to vectorize individual method changes in a way that best represents individual contribution style of each developer, then combine representations of multiple changes made by a single developer into this developer's individual contribution fingerprint.

In the first step, we extract individual changes in Java methods from the project's VCS history and randomly group them into batches of changes authored by the same person. Then we train a neural network to vectorize individual changes and their batches so to distinguish between contributions as efficiently as possible.

At this step, the machine learning model essentially learns a function that maps a code change to a vector. The primary requirement for this function, that defines the learning process, is to represent method changes in a way that groups multiple changes made by one person close to each other and far from changes made by other developers.
In addition, we use an attention mechanism by training the model on batches on multiple code changes instead of single changes. Thanks to attention, the model is also capable of assigning a weight to each method change, defining its ``importance'' for attributing authorship of each change batch.
The inner workings of the authorship attribution model are explained in more detail in Section~\ref{sec:vectorizing}.

In the second step, we combine representations of changes made by every individual developer into a representation for that developer. In this step, we use the trained model for authorship attribution to produce vectors for individual code changes made by the developer, and calculate the representation of a person as a weighed sum of the changes they have made, using attention values from the model as weights. 

In the rest of this section we provide a more detailed technical overview of the extraction pipeline. In the first part, we discuss the inner workings of the authorship attribution model. In the second part, we describe extraction of representations of individual developers from the trained model for authorship recognition.

\begin{figure*}[tp]
\includegraphics[clip,width=\textwidth]{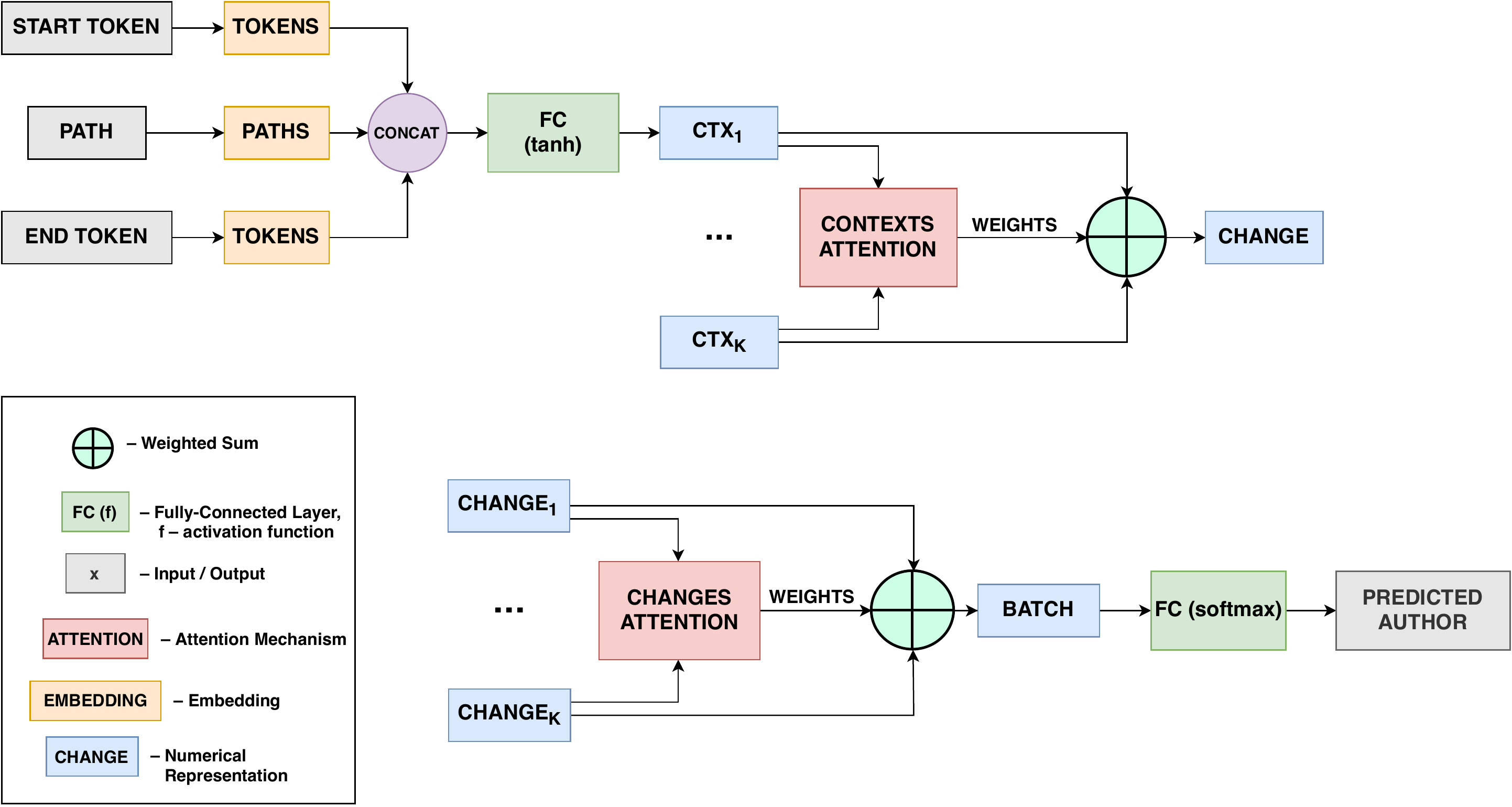}%
\caption{Overview of the authorship attribution pipeline that we use to obtain authorship-based embeddings of method changes and importance of individual method changes for attribution of authorship. The method nodes and their attention weights are later used to produce developer representations.}
\label{fig:method-authorship}
\end{figure*}

\subsection{Vectorizing code changes to represent authorship}\label{sec:vectorizing}
The first step in building the representations of code style is to train a model to distinguish among the contributions of individual developers. During training for authorship attribution, the model implicitly learns to extract information that distinguishes method changes made by each developer from those made by their teammates.

We operate with method changes, rather than static snapshots of code snippets, to use authorship labels from version control: a method change can always be attributed to a single person who performs the change, and this data is already present in the version control system.
Moreover, we use batches of randomly selected method changes as a unit of input for authorship attribution. This allows us to use the whole history of a project for training. For each developer, we shuffle their history of changes and split it into batches. A batch consists of 16 methods authored by the same developer, sampled uniformly randomly from their development history. We chose this number empirically as a good balance point between having too small batches and too little data points per developer. While making the authorship attribution task easier by letting the model focus on more important pieces of input, the use of batches and attention forces the model to estimate importance of specific changes before making a prediction. We further use these attention values when constructing individual style vectors for developers, so to let the code changes that are more representative of a developer contribute more to their fingerprint.

Our authorship recognition model is a neural network. \Cref{fig:method-authorship} presents an overview of the model. Its architecture is based on \textit{code2vec}~\citep{code2vec} -- a state-of-the-art code embedding model. Similarly to \textit{code2vec}, it uses path-based representations~\citep{pbr} of versions of each method before and after a change. 

\subsubsection{Path-based representations.} 
\begin{figure}[htp]
    \centering
    \subfloat[An example code fragment]{
        \includegraphics[clip,width=0.4\columnwidth]{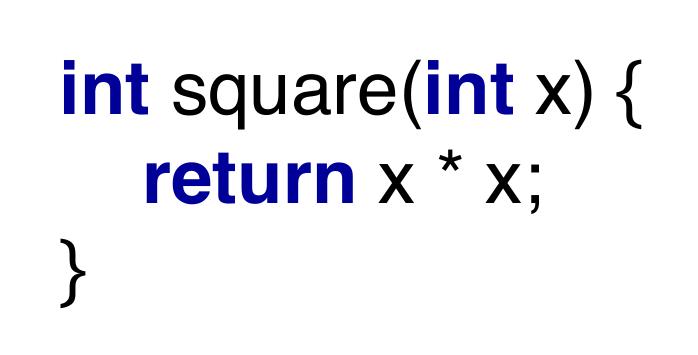}
        \label{fig:ast-code} 
    }
    
    \subfloat[AST of this code fragment]{
       \includegraphics[clip,width=\columnwidth]{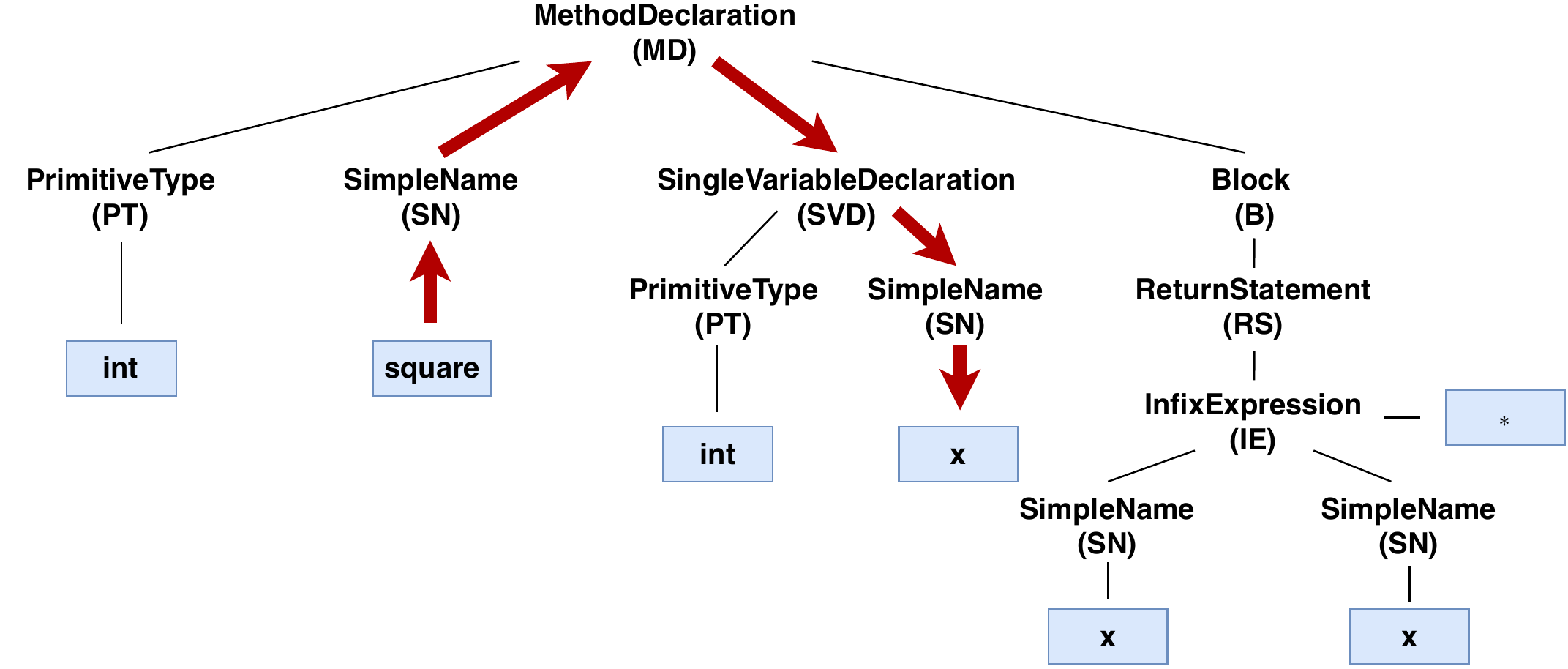}
       \label{fig:ast-example}
    }
    \centering
    \caption{A code example and corresponding AST}
    \label{fig:ast}
\end{figure}

Path-based representations are explained in detail in the original work by Alon et al.~\citep{pbr}. We explain the essential concepts below.\\
\noindent\textbf{Abstract Syntax Tree.}
An abstract syntax tree (AST) is a representation of program's code in the form of a tree. Nodes of the tree correspond to different code constructs (e.g., math operations and variable declarations). Children of a node correspond to smaller constructs that comprise the corresponding code. Different constructs are represented with different \textit{node types}.
An AST omits parentheses, tabs, and other formatting details. \Cref{fig:ast} shows an example of a code fragment and the corresponding AST.\\
\noindent\textbf{AST path.}
A \textit{path} is a sequence of connected nodes in an AST. Start and end nodes of a path may be arbitrary, but we only use paths between two leaves in the AST to conform with code2vec~\cite{code2vec}.
Following Alon et al.~\cite{pbr}, we denote an AST path by a sequence of node types and directions (up or down) between consequent nodes. In \Cref{fig:ast-example}, an example of a path between the leaves of an AST is shown with red arrows. In terms of node types and directions, this path is denoted as follows:
$$
SN \uparrow MD \downarrow SVD \downarrow SN
$$
\noindent\textbf{Path-context.}
The path-based representation operates with \textit{path-contexts}, which are triples consisting of (1) a path between two nodes and the tokens corresponding to (2) start and (3) end nodes. 
From the human perspective, a path-context represents two tokens in code and a structural connection between them. This allows a path-context to capture information about the structure of the code. 
\Cref{fig:ast-example} highlights the following path-context:
$$
(square, SN \uparrow MD \downarrow SVD \downarrow SN, x)
$$

This path-context represents a declaration of a function named \emph{square} with a single argument named \emph{x}. 
The path in this path-context encodes the following information: It contains nodes \textit{Function Declaration} as well as \textit{Single Variable Declaration}. Tokens are linked to \textit{Simple Name} AST nodes.

\subsubsection{Mechanics of the authorship recognition model.}
The first step in the authorship recognition task is to convert a set of individual method changes into a vector form.
For each changed method, we parse both the old method version and the new one to retrieve their ASTs.
We extract path-contexts from both versions of the AST. We impose limitations on maximum length and width of the paths, to only include path-contexts representing local relations in code. 
To distill the concrete effect of each code change, we only use the difference in sets of path-contexts representing the old and the new versions of each method: we only include the path-contexts that were introduced or removed in the new version of the method after the change.
We convert path-contexts representing the difference into a numerical form that can be passed to the neural network. First this,
we apply vocabulary-based encoding to paths and tokens. Vocabulary-based encoding consists in representing every path or token with a unique integer number.

After that we learn the embeddings for tokens and paths. Essentially, at this step the model learns to convert them into a vector form in an optimal way that is most meaningful for the ultimate objective of attributing authorship of method change representations that they comprise. 
Initially represented by a random matrix, stacked embeddings of paths or tokens eventually converge to optimal values during training.

Further down the pipeline, the model concatenates the embedding vector for a path with embeddings of its start and end tokens to build a \textit{path-context vector}. This vector is a combined representation of the path and its end tokens.
We transform the path-context vectors with a fully connected layer and aggregate them into \textit{method change} vectors, using weights from an \emph{attention} layer. The attention mechanism essentially attributes a ``relevance'' weight to each path-context in the batch corresponding to the code change. Path-contexts with higher attention values are more important for distinguishing between developers, i.e., capture more individual information. By highlighting the relevant path-contexts, the attention mechanism improves the accuracy of the model and improves its interpretability: it is possible to pinpoint the concrete path-contexts in the input.

As depicted in the bottom part of Figure \ref{fig:method-authorship}, we use another attention mechanism to combine a batch of method change vectors, each corresponding to a change made to a single method, into a \textit{change batch vector}. Combining changes in batches, rather than using a single change for every prediction, allows to attribute an attention weight, representing its importance for authorship attribution, to each individual method change.
The size of this vector is a hyperparameter of the model, but we choose it to be much less than the number of possible developer labels to ensure that representations of developers are dense.
The next fully connected layer with softmax activation solves the classification problem by learning to attribute a change batch vector to a concrete developer.


\subsection{From authorship recognition to developer embeddings}
The classifier in the authorship recognition model is learning to attribute a change batch vector (which is a weighted sum of change method vectors for methods in the batch) to an individual developer. Essentially, the whole model is learning to map batches of individual method changes into a vector space so that the sets of contributions of individual developers can be separated as well as possible.  High accuracy in the authorship recognition task suggests that the learned model separates the space of method change vectors into areas that correspond to individual developers, thus capturing individual characteristics of a sample of a developer's contributions.

The pivotal idea of our approach is to extract a representation of a developer from the trained model. In addition, the attention mechanism learns to evaluate importance of representations of single method changes in the combined method batch vector.
To extract a representation of contributions of a single developer over a period of time, we consider all of changes made by that developer during the period. We feed the representations of individual changes into the model and retrieve a vector representing each method change as well as the corresponding attention value. Finally, we combine these vectors and weights into a representation of a developer as a weighted sum. 

This representation is used in further analysis: we calculate the representations for multiple time buckets, and retrieve multiple representation vectors for various team members, each corresponding to a certain time period. Finally, we explore whether positions and relative movement of representations are connected to learning from peers, as reported by developers.

\begin{figure}[htp]
\includegraphics[clip,width=\columnwidth]{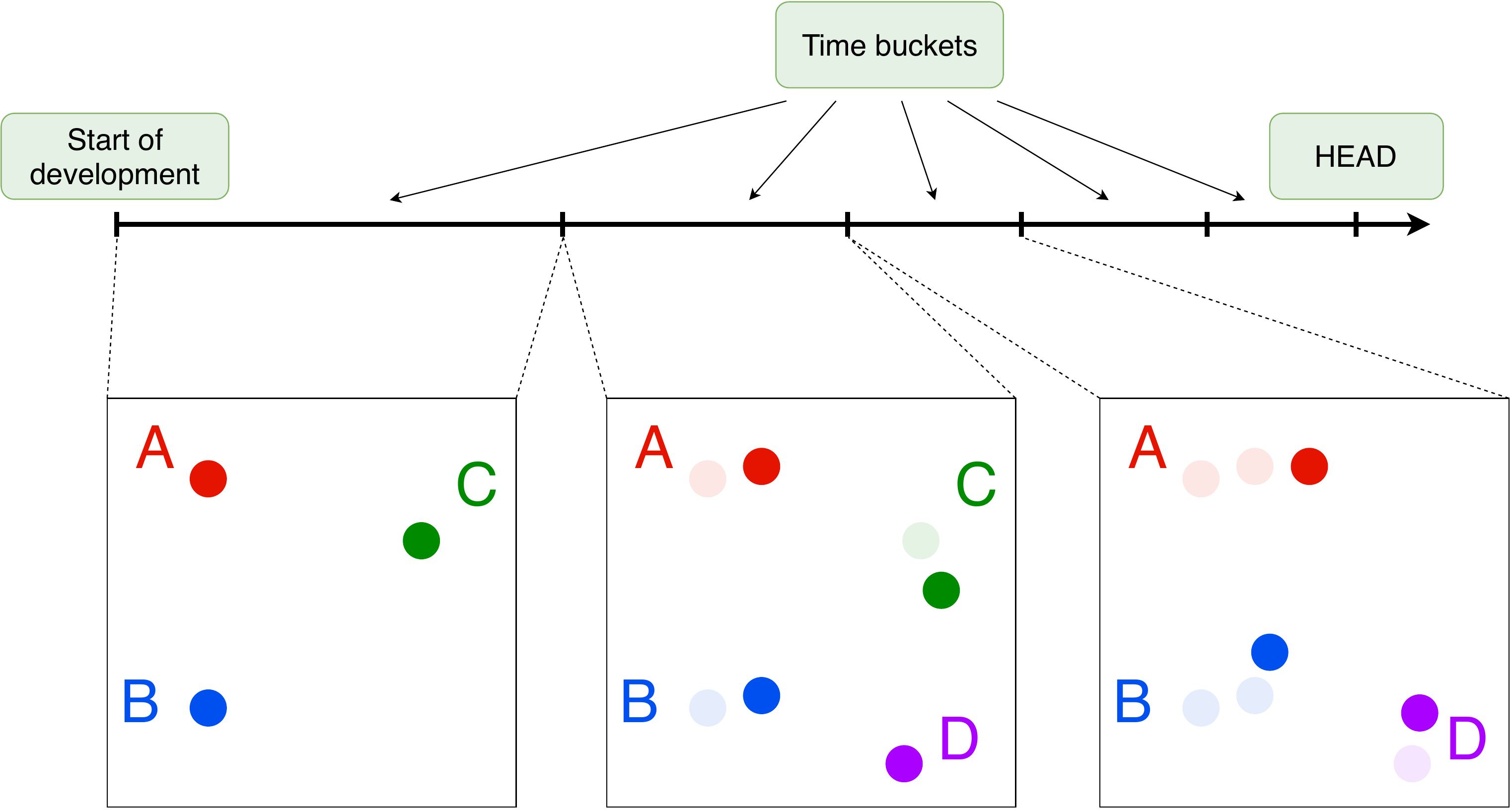}%
\caption{Change of developers' vectors between time buckets. A developer may be only active in some of the buckets.}
\label{fig:codestyle-time}
\end{figure}

\subsection{Threats to validity}
\textbf{The curse of context.}
The authorship recognition model distinguishes contributors based on both the \textit{structure} of code (which is represented by sequences of node types from AST paths) and the \textit{context} of their changes (which is represented by the tokens). Tokens include variable names and names of declared and invoked methods. These names may be highly specific for a concrete narrow area of code. It is reasonable to think that in projects with practices of individual code ownership this context information alone can be enough to recognize the author of a method change. 
Given our ultimate goal of capturing individual characteristics of developers, including the context information in the model is not always desirable.
While we perform a separate evaluation on data with excluded tokens, there is a chance that context information may as well be reflected in characteristic unique code patterns that are captured in sequences of AST node types.\smallskip
\textbf{Performance.}
We must note that resource consumption of our approach is very high, mostly due to the need to repeat training multiple times to reduce noise in the data. While not a crucial property for a proof-of-concept tool, reasonable performance of an approach is a requirement for practical applicability of an approach to this task. Producing a slow pipeline for a far-fetched, yet practical, goal impacts strength of the motivation.

\section{Evaluation Setup}
One critical design choice for our approach to building individual developer representations, or embeddings, is to take a step away of explicitly defined code style. In essence, as we use the authorship attribution task for building the representations, we implicitly define code style as \emph{``anything that distinguishes between individuals' code''}. While avoiding explicit definition of code style gives potential to include characteristics of the style that are otherwise left out, this makes the evaluation of embeddings' quality challenging, as there is no ground truth to compare against.

To get a realistic estimate of quality of code style embeddings, we decided to evaluate them in the context of a possible application. 
As discussed in Section~\ref{sec:representations-need}, one promising application of code style representations is enabling team collaboration tools to make sense of proximity of individual contribution styles, capture the process of knowledge transfer in teams, and potentially provide aid by aligning this process to be more efficient. 
Essentially, we formulate the task of embeddings evaluation as evaluating their ability to \emph{capture learning between individuals}. In the rest of this section, we elaborate on the evaluation setup and technique.

\subsection{Dataset preparation}
As an evaluation dataset, we use the source code and development history of IntelliJ Community.\footnote{\url{https://github.com/jetbrains/intellij-community}}\smallskip

\textbf{Merging, splitting and filtering.}
In the first step, we merge name-email pairs accounted in the VCS history and belonging to the same developer into a single entity. For this purpose, we used a separate user management tool, internally used by developers of IntelliJ and accessible to us, containing merged records of VCS entities for developers in the project. To facilitate running our pipeline on other projects, we implemented a simple algorithm for entities merging: it builds a bipartite graph of names and emails, where pairs that appear together are connected by an edge. Connected components in the graph correspond to merged entities. While the algorithm is not perfect, it gives an approximate merging which can be further improved manually. We include the implementation into the reproduction package.

Afterward, we split the history of the repository into multiple time chunks, each containing the same number of commits. Learning representations of developers over a small time chunk, rather than over complete history of a repository, allows to produce multiple representations of a developer, each corresponding to a relatively short time bucket. This accounts for potential changes in developers' coding traits over time and allows us to look at changes in distances between representations in consecutive time buckets -- in other words, to track relative movement of representations.

As explained in Section~\ref{sec:method}, we focus on Java method changes for training and producing the representations. 
The repository contains contributions from about 500 developers, with a long tail of developers with only a few contributions. To ensure that we have sufficient data for every developer, we exclude the developers who made less than 1,000 method changes over the whole history of the repository. This left us with 124 active developers in the dataset. Having significant amounts of data for every developer, we increase stability of representations and reduce noise.\smallskip

\textbf{Noise reduction.}
The weights in the authorship attribution model, which ultimately represent the resulting change embedding function, are randomly initialized before learning and may converge to very different configurations depending on the initial random seed. Moreover, density of representations in each snapshot for multiple developers differs.

To account for the varying density, we calculate (for each time bucket) the average distance between every pair of developer representations, and divide the actual distances by this value. This allows us to compare distances between two given representations in consecutive time buckets and is necessary because density of representations may differ between two buckets. 

On top of normalization, to account for the random nature of representations, we repeat the whole learning and style representation extraction 30 times and calculate the average normalized distance between every two developers in every time bucket across all runs. While making the process of obtaining representations from a large repository computationally demanding, repeating the learning and using the average make the resulting data less noisy and the comparison of distances between two different pairs more reliable. 

The resulting data consists of 20 lists of relative distances between every two representations, with each snapshot corresponding to a certain time bucket.
The number of resulting representations in a bucket varies between 23 and 87, depending on the bucket, displaying an upward trend, representing the growth of the team.


\subsection{Team survey}
To get a baseline to check to what extent proximity and relative movement of individual representations, as extracted by the model, reflect actual learning in the team,
we circulated a short online survey with the development team of the project in the dataset via a post in the project's internal communication channel.

\begin{figure}[htp]
\frame{
    \includegraphics[clip,width=\columnwidth]{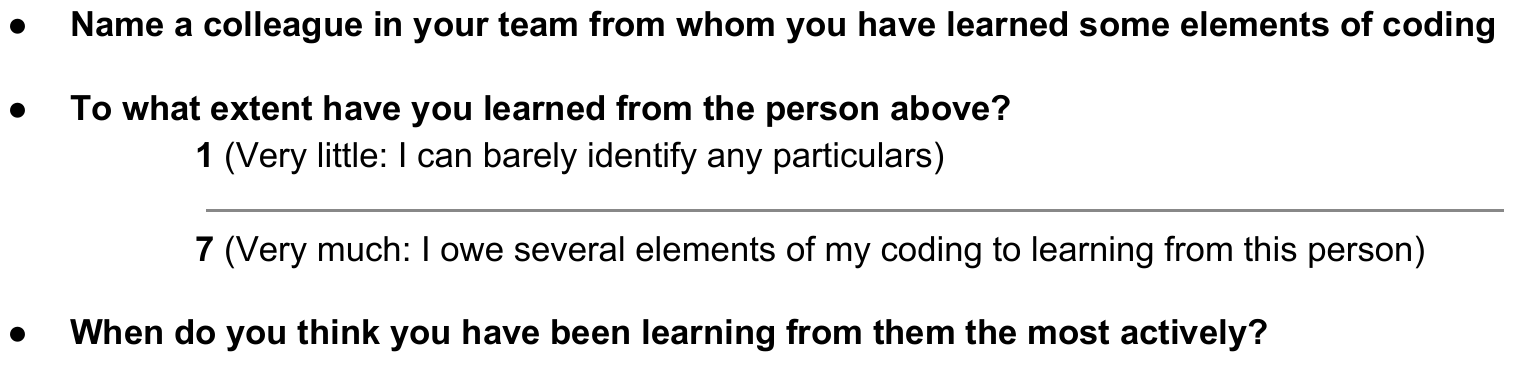}
}%
\caption{Excerpt from the survey}
\label{fig:survey}
\end{figure}

The goal of the survey was to collect information on mutual learning between some developers in the team, so that this data could be used as ground truth regarding actual learning taking place, which could let us see whether relative movement of developer representation reflected learning reported by developers. 
In the survey, we ask each respondents the questions depicted in \Cref{fig:survey} three times to get information about three different colleagues they have learned from. We also included an option to mention one or two colleagues in a similar way. To avoid hinting the respondents with any particular definition of coding elements their learning may relate to, we explicitly stated that we would like them to define it for themselves and consider anything they may have learned during collaborative work, mentoring, code review, or other team activities. Responses to this question provide us with examples of positive pairs in terms of reported learning.

In addition, we ask the respondents to mention several developers who contribute to the same project but from whom they are sure they have not learned any elements of coding:
\textit{``Please name a few colleagues from the IntelliJ team you think you did not learn any coding elements from at all''}.
This question provides us with a set of negative examples.

\subsection{Survey results and model output}
The actual evaluation consisted in mapping the results of the developer survey onto data of relative movement of developers' representations between different time buckets.

While we asked the survey participants to indicate the degree of perceived learning and the time period when such a learning took place, we believe that the data for these two questions is not reliable enough. 3 respondents simply reported a similar neutral or extreme score for every person they mention and noted that they were confused by the request to indicate the degree of learning. Regarding the time period when learning took place, only one in three respondents provided any meaningful information, which in most cases was still too vague to attribute to a concrete time bucket.
Thus, we only use the fact of a participant naming someone they have learned from, or certainly have not learned from, as extracted from the answers, to map the results of the survey onto the output of the model. 
The final data from the survey consists of 23 \emph{positive pairs} -- pairs of developers one of whom reporting learning from the other, and 13 \emph{negative pairs}, where one of the developers named the other as someone they certainly have not learned any elements of coding from. 

To see whether distances between representations reflect reported learning, we compare distributions of distances in all positive pairs, taken across all buckets where both developers in the pair displayed activity and were present in the data, to a similarly defined distribution across all negative pairs. The sample of distances for positive pairs consists of 229 values, and a similar sample for negative pairs has 113 values.

To extract dynamics of relative distances, we obtain a distribution of \emph{differences of distances between two consecutive time buckets} for all pairs in the positive and negative groups, using every time bucket where these distances are defined both in the given and the previous bucket. The sample of distance differences for positive pairs contains 204 values, and the negative pairs yielded 99 values. 

Finally, we use a 2-sample Kolmogorov-Smirnov test\footnote{We use this test because we cannot make any assumptions about distribution of these values due to the obscure stochastic process of learning and the fact that this data may reflect some social processes in teams that are impossible to completely quantify.} to compare distributions of distances, as well as distributions of their differences, between the positive and negative groups.

\section{Evaluation results}

\textbf{Relative distances.}
%
When tokens are present in the input data, the distribution of 229 distance values in pairs with reported learning has a mean of $0.85$ and variance of $0.54$. For pairs with reported lack of learning, 113 distance values display a mean of $1.19$ with variance of $0.60$. The p-value from the 2-sample Kolmogorov-Smirnov test comparing distributions of these two samples is under $10^{-7}$.
When tokens are removed from the input data (to minimize context information available to the model), the mean distance between positive pairs is $0.81$ with a variance of $1.37$. For negative pairs the mean distance is $0.98$ and variance is $1.22$. The KS test yields a p-value $< 0.001$.

These results suggest that the \emph{representations of developers in pairs with reported learning are located closer to each other, both in cases with included and excluded tokens.}\smallskip

\textbf{Relative movement.}
We perform a similar comparison for the distributions of distances in pairs of representations between two consequent time buckets.
With tokens in data, 204 values for positive pairs are distributed with mean of $-0.005$ and variance of $0.86$. 99 values for negative pairs display a mean of $0.026$ and variance of $0.59$. The KS test yields a p-value of $0.94$, suggesting that samples of distance differences for positive and negative pairs are likely from similar distributions.
When no tokens are present, values for positive pairs display a distribution with mean of $0.046$ and variance of $1.97$. Values for negative pairs are distributed with mean of $-0.045$ and variance of $2.47$. The p-value in the KS test is 0.16.

These results suggest that \emph{distributions of differences in relative distances between consecutive time buckets are no different between pairs with reported learning and with reported lack of learning, regardless of whether tokens are included.}\smallskip

\textbf{Summary.} Overall, learning is to an extent reflected in the distances between representations: distances between developers who report learning from each other are lower than between developers who report not learning from each other. This result persists when tokens are removed from the data, thus importantly suggesting how learning is not only captured in context of developers' contributions represented by tokens.
However, a similar comparison for distributions of distance differences between consecutive time buckets suggests that distributions are similar for both groups of developer pairs.
\section{Discussion}
The core idea of our approach to building coding style representations is to step aside from explicitly defined feature sets for developer representations, rather build the representations of style by aggregating embeddings of code changes, which are produced to distinguish between developers as good as possible by the authorship attribution model.

We evaluated the computed developer embeddings through reports of peer learning in a large development team finding that inter-peer learning is indeed reflected in the embeddings: representations of developers who a given developer reports having learned from are closer to their own representation than representations of those who they reported not having learned from. Importantly, this effect persists even when the context information (reflected in concrete code tokens) is removed from the training data.
This result suggests that implicitly built developer representations reflect the fuzzy process of learning in teams, by capturing individual characteristic patterns of code constructs and their proximity for people who report learning from each other and not just capturing the context information.

However, we do not see any connection between reported learning and relative movement of developer representations. In our opinion, the most important reason for the lack of such a connection is the low stability of the resulting representations. While reflecting learning in distributions of relative positions of individual representations, the representations are too noisy to reason about learning in side-by-side comparison of team snapshots in consecutive time buckets. We discuss possible ways to mitigate this in the next subsection.

\subsection{Future work}\label{future-work}



\textbf{Evaluation.}
A more elaborate evaluation of our approach, involving multiple software projects and more feedback from developers, could help clarify quality of the representations.\\
\textbf{Other scopes.}
Apart from the learning detection task, the representations produced by our approach could find use in other contexts, including tasks already supported by team collaboration tools, e.g., recommendation of code reviewers~\cite{kovalenko2018does} or issue assignment~\cite{anvik2006should}. \\
\textbf{Alternative embedding techniques.}
We use a modified version of code2vec for method change embeddings. Using other embedding techniques could improve the performance of our approach.\\
\textbf{Transparency.}
The authorship attribution model uses two attention layers: one to learn importance of individual changes in a batch and the other for individual path-contexts. While we use weights from the former to produce developer representations, a careful consideration of values from both layers could provide more insights on what makes a certain change characteristic for a developer.\\
\textbf{Stability of representations.}
Noise and jitter in representation snapshots between time buckets make extracting representations harder. Additional constraints could help increase representations' stability. Figuring out a way to increase stability without compromising ability of representations to capture important social properties is another promising direction of future work.

\section{Conclusion}
We introduced an approach to building representations of individual coding style of developers relative to their peers in the team. The most important feature of our approach is that it does not require explicit feature engineering, rather relies on implicit vectorization of code changes via a code embedding model, trained to distinguish between changes made by individual developers and the aggregation of individual changes.

We demonstrate that it is possible to build representation of individual developers' coding style without defining style formally. The resulting representations reflect learning between peers in the team to a certain degree.\smallskip

\textbf{Reproducibility.} The technical artifacts of this work -- the pipeline to build the representations and data analysis code that we used to map the survey results to the data -- are available online: \url{https://zenodo.org/record/3647645}

\bibliographystyle{ACM-Reference-Format}
\bibliography{codestyle-ml4se}


\begin{thebibliography}{00}


\ifx \showCODEN    \undefined \def \showCODEN     #1{\unskip}     \fi
\ifx \showDOI      \undefined \def \showDOI       #1{#1}\fi
\ifx \showISBNx    \undefined \def \showISBNx     #1{\unskip}     \fi
\ifx \showISBNxiii \undefined \def \showISBNxiii  #1{\unskip}     \fi
\ifx \showISSN     \undefined \def \showISSN      #1{\unskip}     \fi
\ifx \showLCCN     \undefined \def \showLCCN      #1{\unskip}     \fi
\ifx \shownote     \undefined \def \shownote      #1{#1}          \fi
\ifx \showarticletitle \undefined \def \showarticletitle #1{#1}   \fi
\ifx \showURL      \undefined \def \showURL       {\relax}        \fi
\providecommand\bibfield[2]{#2}
\providecommand\bibinfo[2]{#2}
\providecommand\natexlab[1]{#1}
\providecommand\showeprint[2][]{arXiv:#2}

\bibitem[\protect\citeauthoryear{??}{ecl}{2019a}]%
        {eclipse-cdt}
 \bibinfo{year}{2019}\natexlab{a}.
\newblock \bibinfo{title}{Eclipse CDT | The Eclipse Foundation}.
\newblock \bibinfo{howpublished}{\url{https://www.eclipse.org/cdt/}}.
  (\bibinfo{year}{2019}).
\newblock
\newblock
\shownote{Accessed: 2019-12-27.}


\bibitem[\protect\citeauthoryear{??}{ecl}{2019b}]%
        {eclipse-ide}
 \bibinfo{year}{2019}\natexlab{b}.
\newblock \bibinfo{title}{Eclipse desktop \& web IDEs | The Eclipse
  Foundation}.
\newblock \bibinfo{howpublished}{\url{https://www.eclipse.org/ide/}}.
  (\bibinfo{year}{2019}).
\newblock
\newblock
\shownote{Accessed: 2019-12-27.}


\bibitem[\protect\citeauthoryear{??}{ecl}{2019c}]%
        {eclipse-jdt}
 \bibinfo{year}{2019}\natexlab{c}.
\newblock \bibinfo{title}{Eclipse Java development tools (JDT) | The Eclipse
  Foundation}.
\newblock \bibinfo{howpublished}{\url{https://www.eclipse.org/jdt/}}.
  (\bibinfo{year}{2019}).
\newblock
\newblock
\shownote{Accessed: 2019-12-27.}


\bibitem[\protect\citeauthoryear{??}{int}{2019a}]%
        {intellij-idea}
 \bibinfo{year}{2019}\natexlab{a}.
\newblock \bibinfo{title}{IntelliJ IDEA: The Java IDE for Professional
  Developers by JetBrains}.
\newblock \bibinfo{howpublished}{\url{https://www.jetbrains.com/idea/}}.
  (\bibinfo{year}{2019}).
\newblock
\newblock
\shownote{Accessed: 2019-12-27.}


\bibitem[\protect\citeauthoryear{??}{int}{2019b}]%
        {intellij-psi}
 \bibinfo{year}{2019}\natexlab{b}.
\newblock \bibinfo{title}{Program Structure Interface (PSI)}.
\newblock
  \bibinfo{howpublished}{\url{https://www.jetbrains.org/intellij/sdk/docs/basics/architectural_overview/psi.html}}.
    (\bibinfo{year}{2019}).
\newblock
\newblock
\shownote{Accessed: 2019-12-26.}


\bibitem[\protect\citeauthoryear{Allamanis, Peng, and Sutton}{Allamanis
  et~al\mbox{.}}{2016}]%
        {allamanis2016convolutional}
\bibfield{author}{\bibinfo{person}{Miltiadis Allamanis}, \bibinfo{person}{Hao
  Peng}, {and} \bibinfo{person}{Charles Sutton}.}
  \bibinfo{year}{2016}\natexlab{}.
\newblock \showarticletitle{A convolutional attention network for extreme
  summarization of source code}. In \bibinfo{booktitle}{{\em International
  Conference on Machine Learning}}. \bibinfo{pages}{2091--2100}.
\newblock


\bibitem[\protect\citeauthoryear{Alon, Zilberstein, Levy, and Yahav}{Alon
  et~al\mbox{.}}{2018}]%
        {pbr}
\bibfield{author}{\bibinfo{person}{Uri Alon}, \bibinfo{person}{Meital
  Zilberstein}, \bibinfo{person}{Omer Levy}, {and} \bibinfo{person}{Eran
  Yahav}.} \bibinfo{year}{2018}\natexlab{}.
\newblock \showarticletitle{A general path-based representation for predicting
  program properties}. In \bibinfo{booktitle}{{\em Proceedings of the 39th ACM
  SIGPLAN Conference on Programming Language Design and Implementation}}. ACM,
  \bibinfo{pages}{404--419}.
\newblock


\bibitem[\protect\citeauthoryear{Alon, Zilberstein, Levy, and Yahav}{Alon
  et~al\mbox{.}}{2019}]%
        {code2vec}
\bibfield{author}{\bibinfo{person}{Uri Alon}, \bibinfo{person}{Meital
  Zilberstein}, \bibinfo{person}{Omer Levy}, {and} \bibinfo{person}{Eran
  Yahav}.} \bibinfo{year}{2019}\natexlab{}.
\newblock \showarticletitle{code2vec: Learning distributed representations of
  code}.
\newblock \bibinfo{journal}{{\em Proceedings of the ACM on Programming
  Languages\/}} \bibinfo{volume}{3}, \bibinfo{number}{POPL}
  (\bibinfo{year}{2019}), \bibinfo{pages}{40}.
\newblock


\bibitem[\protect\citeauthoryear{Alsulami, Dauber, Harang, Mancoridis, and
  Greenstadt}{Alsulami et~al\mbox{.}}{2017}]%
        {alsulami2017source}
\bibfield{author}{\bibinfo{person}{Bander Alsulami}, \bibinfo{person}{Edwin
  Dauber}, \bibinfo{person}{Richard Harang}, \bibinfo{person}{Spiros
  Mancoridis}, {and} \bibinfo{person}{Rachel Greenstadt}.}
  \bibinfo{year}{2017}\natexlab{}.
\newblock \showarticletitle{Source code authorship attribution using long
  short-term memory based networks}. In \bibinfo{booktitle}{{\em European
  Symposium on Research in Computer Security}}. Springer,
  \bibinfo{pages}{65--82}.
\newblock


\bibitem[\protect\citeauthoryear{Anvik, Hiew, and Murphy}{Anvik
  et~al\mbox{.}}{2006}]%
        {anvik2006should}
\bibfield{author}{\bibinfo{person}{John Anvik}, \bibinfo{person}{Lyndon Hiew},
  {and} \bibinfo{person}{Gail~C Murphy}.} \bibinfo{year}{2006}\natexlab{}.
\newblock \showarticletitle{Who should fix this bug?}. In
  \bibinfo{booktitle}{{\em Proceedings of the 28th international conference on
  Software engineering}}. \bibinfo{pages}{361--370}.
\newblock


\bibitem[\protect\citeauthoryear{Azcona, Arora, Hsiao, and Smeaton}{Azcona
  et~al\mbox{.}}{2019}]%
        {azcona2019user2code2vec}
\bibfield{author}{\bibinfo{person}{David Azcona}, \bibinfo{person}{Piyush
  Arora}, \bibinfo{person}{I-Han Hsiao}, {and} \bibinfo{person}{Alan Smeaton}.}
  \bibinfo{year}{2019}\natexlab{}.
\newblock \showarticletitle{user2code2vec: Embeddings for Profiling Students
  Based on Distributional Representations of Source Code}. In
  \bibinfo{booktitle}{{\em Proceedings of the 9th International Conference on
  Learning Analytics \& Knowledge}}. ACM, \bibinfo{pages}{86--95}.
\newblock


\bibitem[\protect\citeauthoryear{Bacchelli and Bird}{Bacchelli and
  Bird}{2013}]%
        {bacchelli2013expectations}
\bibfield{author}{\bibinfo{person}{Alberto Bacchelli} {and}
  \bibinfo{person}{Christian Bird}.} \bibinfo{year}{2013}\natexlab{}.
\newblock \showarticletitle{Expectations, outcomes, and challenges of modern
  code review}. In \bibinfo{booktitle}{{\em Proceedings of the 2013
  international conference on software engineering}}. IEEE Press,
  \bibinfo{pages}{712--721}.
\newblock


\bibitem[\protect\citeauthoryear{Bahdanau, Cho, and Bengio}{Bahdanau
  et~al\mbox{.}}{2014}]%
        {bahdanau2014neural}
\bibfield{author}{\bibinfo{person}{Dzmitry Bahdanau},
  \bibinfo{person}{Kyunghyun Cho}, {and} \bibinfo{person}{Yoshua Bengio}.}
  \bibinfo{year}{2014}\natexlab{}.
\newblock \showarticletitle{Neural machine translation by jointly learning to
  align and translate}.
\newblock \bibinfo{journal}{{\em arXiv preprint arXiv:1409.0473\/}}
  (\bibinfo{year}{2014}).
\newblock


\bibitem[\protect\citeauthoryear{Caliskan-Islam, Harang, Liu, Narayanan, Voss,
  Yamaguchi, and Greenstadt}{Caliskan-Islam et~al\mbox{.}}{2015}]%
        {caliskan2015deanonymizing}
\bibfield{author}{\bibinfo{person}{Aylin Caliskan-Islam},
  \bibinfo{person}{Richard Harang}, \bibinfo{person}{Andrew Liu},
  \bibinfo{person}{Arvind Narayanan}, \bibinfo{person}{Clare Voss},
  \bibinfo{person}{Fabian Yamaguchi}, {and} \bibinfo{person}{Rachel
  Greenstadt}.} \bibinfo{year}{2015}\natexlab{}.
\newblock \showarticletitle{De-anonymizing Programmers via Code Stylometry}. In
  \bibinfo{booktitle}{{\em 24th {USENIX} Security Symposium ({USENIX} Security
  15)}}. \bibinfo{publisher}{{USENIX} Association},
  \bibinfo{address}{Washington, D.C.}, \bibinfo{pages}{255--270}.
\newblock
\showISBNx{978-1-931971-232}
\showURL{%
\url{https://www.usenix.org/conference/usenixsecurity15/technical-sessions/presentation/caliskan-islam}}


\bibitem[\protect\citeauthoryear{Cataldo, Herbsleb, and Carley}{Cataldo
  et~al\mbox{.}}{2008}]%
        {cataldo2008socio}
\bibfield{author}{\bibinfo{person}{Marcelo Cataldo}, \bibinfo{person}{James~D
  Herbsleb}, {and} \bibinfo{person}{Kathleen~M Carley}.}
  \bibinfo{year}{2008}\natexlab{}.
\newblock \showarticletitle{Socio-technical congruence: a framework for
  assessing the impact of technical and work dependencies on software
  development productivity}. In \bibinfo{booktitle}{{\em Proceedings of the
  Second ACM-IEEE international symposium on Empirical software engineering and
  measurement}}. ACM, \bibinfo{pages}{2--11}.
\newblock


\bibitem[\protect\citeauthoryear{Devlin, Uesato, Bhupatiraju, Singh, Mohamed,
  and Kohli}{Devlin et~al\mbox{.}}{2017}]%
        {devlin2017robustfill}
\bibfield{author}{\bibinfo{person}{Jacob Devlin}, \bibinfo{person}{Jonathan
  Uesato}, \bibinfo{person}{Surya Bhupatiraju}, \bibinfo{person}{Rishabh
  Singh}, \bibinfo{person}{Abdel-rahman Mohamed}, {and}
  \bibinfo{person}{Pushmeet Kohli}.} \bibinfo{year}{2017}\natexlab{}.
\newblock \showarticletitle{Robustfill: Neural program learning under noisy
  I/O}. In \bibinfo{booktitle}{{\em Proceedings of the 34th International
  Conference on Machine Learning-Volume 70}}. JMLR. org,
  \bibinfo{pages}{990--998}.
\newblock


\bibitem[\protect\citeauthoryear{Falleri, Morandat, Blanc, Martinez, and
  Monperrus}{Falleri et~al\mbox{.}}{2014}]%
        {DBLP:conf/kbse/FalleriMBMM14}
\bibfield{author}{\bibinfo{person}{Jean{-}R{\'{e}}my Falleri},
  \bibinfo{person}{Flor{\'{e}}al Morandat}, \bibinfo{person}{Xavier Blanc},
  \bibinfo{person}{Matias Martinez}, {and} \bibinfo{person}{Martin Monperrus}.}
  \bibinfo{year}{2014}\natexlab{}.
\newblock \showarticletitle{Fine-grained and accurate source code
  differencing}. In \bibinfo{booktitle}{{\em {ACM/IEEE} International
  Conference on Automated Software Engineering, {ASE} '14, Vasteras, Sweden -
  September 15 - 19, 2014}}. \bibinfo{pages}{313--324}.
\newblock
\showDOI{%
\url{https://doi.org/10.1145/2642937.2642982}}


\bibitem[\protect\citeauthoryear{Haiduc, Aponte, and Marcus}{Haiduc
  et~al\mbox{.}}{2010}]%
        {haiduc2010supporting}
\bibfield{author}{\bibinfo{person}{Sonia Haiduc}, \bibinfo{person}{Jairo
  Aponte}, {and} \bibinfo{person}{Andrian Marcus}.}
  \bibinfo{year}{2010}\natexlab{}.
\newblock \showarticletitle{Supporting program comprehension with source code
  summarization}. In \bibinfo{booktitle}{{\em Proceedings of the 32nd ACM/IEEE
  International Conference on Software Engineering-Volume 2}}. ACM,
  \bibinfo{pages}{223--226}.
\newblock


\bibitem[\protect\citeauthoryear{Kovalenko, Tintarev, Pasynkov, Bird, and
  Bacchelli}{Kovalenko et~al\mbox{.}}{2018}]%
        {kovalenko2018does}
\bibfield{author}{\bibinfo{person}{Vladimir Kovalenko}, \bibinfo{person}{Nava
  Tintarev}, \bibinfo{person}{Evgeny Pasynkov}, \bibinfo{person}{Christian
  Bird}, {and} \bibinfo{person}{Alberto Bacchelli}.}
  \bibinfo{year}{2018}\natexlab{}.
\newblock \showarticletitle{Does reviewer recommendation help developers?}
\newblock \bibinfo{journal}{{\em IEEE Transactions on Software Engineering\/}}
  (\bibinfo{year}{2018}).
\newblock


\bibitem[\protect\citeauthoryear{Lange and Mancoridis}{Lange and
  Mancoridis}{2007}]%
        {lange2007using}
\bibfield{author}{\bibinfo{person}{Robert~Charles Lange} {and}
  \bibinfo{person}{Spiros Mancoridis}.} \bibinfo{year}{2007}\natexlab{}.
\newblock \showarticletitle{Using code metric histograms and genetic algorithms
  to perform author identification for software forensics}. In
  \bibinfo{booktitle}{{\em Proceedings of the 9th annual conference on Genetic
  and evolutionary computation}}. \bibinfo{pages}{2082--2089}.
\newblock


\bibitem[\protect\citeauthoryear{Raychev, Vechev, and Yahav}{Raychev
  et~al\mbox{.}}{2014}]%
        {raychev2014code}
\bibfield{author}{\bibinfo{person}{Veselin Raychev}, \bibinfo{person}{Martin
  Vechev}, {and} \bibinfo{person}{Eran Yahav}.}
  \bibinfo{year}{2014}\natexlab{}.
\newblock \showarticletitle{Code Completion with Statistical Language Models}.
  In \bibinfo{booktitle}{{\em Proceedings of the 35th ACM SIGPLAN Conference on
  Programming Language Design and Implementation}} {\em (\bibinfo{series}{PLDI
  '14})}. \bibinfo{publisher}{ACM}, \bibinfo{address}{New York, NY, USA},
  \bibinfo{pages}{419--428}.
\newblock
\showISBNx{978-1-4503-2784-8}
\showDOI{%
\url{https://doi.org/10.1145/2594291.2594321}}


\bibitem[\protect\citeauthoryear{Thongtanunam, Tantithamthavorn, Kula, Yoshida,
  Iida, and Matsumoto}{Thongtanunam et~al\mbox{.}}{2015}]%
        {thongtanunam2015should}
\bibfield{author}{\bibinfo{person}{Patanamon Thongtanunam},
  \bibinfo{person}{Chakkrit Tantithamthavorn}, \bibinfo{person}{Raula~Gaikovina
  Kula}, \bibinfo{person}{Norihiro Yoshida}, \bibinfo{person}{Hajimu Iida},
  {and} \bibinfo{person}{Ken-ichi Matsumoto}.} \bibinfo{year}{2015}\natexlab{}.
\newblock \showarticletitle{Who should review my code? a file location-based
  code-reviewer recommendation approach for modern code review}. In
  \bibinfo{booktitle}{{\em 2015 IEEE 22nd International Conference on Software
  Analysis, Evolution, and Reengineering (SANER)}}. IEEE,
  \bibinfo{pages}{141--150}.
\newblock


\bibitem[\protect\citeauthoryear{Tran, Chulkov, and Sch{\"o}nw{\"a}lder}{Tran
  et~al\mbox{.}}{2008}]%
        {tran2008crawling}
\bibfield{author}{\bibinfo{person}{Ha~Manh Tran}, \bibinfo{person}{Georgi
  Chulkov}, {and} \bibinfo{person}{J{\"u}rgen Sch{\"o}nw{\"a}lder}.}
  \bibinfo{year}{2008}\natexlab{}.
\newblock \showarticletitle{Crawling bug tracker for semantic bug search}. In
  \bibinfo{booktitle}{{\em International Workshop on Distributed Systems:
  Operations and Management}}. Springer, \bibinfo{pages}{55--68}.
\newblock


\bibitem[\protect\citeauthoryear{Valetto, Helander, Ehrlich, Chulani, Wegman,
  and Williams}{Valetto et~al\mbox{.}}{2007}]%
        {valetto2007using}
\bibfield{author}{\bibinfo{person}{Giuseppe Valetto}, \bibinfo{person}{Mary
  Helander}, \bibinfo{person}{Kate Ehrlich}, \bibinfo{person}{Sunita Chulani},
  \bibinfo{person}{Mark Wegman}, {and} \bibinfo{person}{Clay Williams}.}
  \bibinfo{year}{2007}\natexlab{}.
\newblock \showarticletitle{Using software repositories to investigate
  socio-technical congruence in development projects}. In
  \bibinfo{booktitle}{{\em Fourth International Workshop on Mining Software
  Repositories (MSR'07: ICSE Workshops 2007)}}. IEEE, \bibinfo{pages}{25--25}.
\newblock


\end{thebibliography}

\end{document}